\documentclass[aps,prb,twocolumn,showpacs,superscriptaddress]{revtex4}

\usepackage{graphicx} 
\usepackage{multirow}
\usepackage[T1]{fontenc}
\usepackage[latin9]{inputenc}
\usepackage{amsmath}
\usepackage{amssymb}

\date{\today}
\begin{document}
\title{Graphene coatings: An efficient protection from oxidation}

\author{M. Topsakal}
\affiliation{UNAM-National Nanotechnology Research Center, Bilkent University, 06800 Ankara, Turkey}
\affiliation{Institute of Materials Science and Nanotechnology, Bilkent University, 06800 Ankara, Turkey}
\author{H. \c{S}ahin}
\affiliation{UNAM-National Nanotechnology Research Center, Bilkent University, 06800 Ankara, Turkey}
\affiliation{Institute of Materials Science and Nanotechnology, Bilkent University, 06800 Ankara, Turkey}
\author{S. Ciraci}\email{ciraci@fen.bilkent.edu.tr}
\affiliation{UNAM-National Nanotechnology Research Center, Bilkent University, 06800 Ankara, Turkey}
\affiliation{Institute of Materials Science and Nanotechnology, Bilkent University, 06800 Ankara, Turkey}
\affiliation{Department of Physics, Bilkent University,
06800 Ankara, Turkey}

\date{\today}

\begin{abstract}
We demonstrate that graphene coating can provide an efficient protection from
oxidation by posing a high energy barrier to the path of oxygen atom, which could
have penetrated from the top of graphene to the reactive surface underneath. Graphene bilayer,
which blocks the diffusion of oxygen with a relatively higher energy barrier provides
even better protection from oxidation. While an oxygen molecule is weakly bound to bare
graphene surface and hence becomes rather inactive, it can easily dissociates
into two oxygen atoms adsorbed to low coordinated carbon atoms at the edges of a vacancy.
For these oxygen atoms the oxidation barrier is reduced and hence the protection from
oxidation provided by graphene coatings is weakened. Our predictions obtained from the
state of the art first-principles calculations of electronic structure, phonon density
of states and reaction path will unravel how a graphene can be used as a corrosion
resistant coating and guide further studies aiming at developing more efficient nanocoatings.
\end{abstract}

\maketitle

\section{Introduction}
The reaction of material surfaces with oxygen and controlling damages from
corrosion have been the subject of intensive research for decades. While
protective macroscale coatings give rise to the modification of the sizes and some
other physical properties of reactive surfaces, progress made to date has revealed
several advantages of nanoscale coatings in protection from corrosion and wear.

The earliest efforts of protection from oxidation using carbon based materials were 
devoted to the carbon deposition on metal surfaces. It was reported that Ni and Co 
surfaces can be covered by carbon deposition at high temperatures.\cite{derbyshire} 
It was shown that the (111) surfaces of the Ni single crystals can be covered by monolayer 
carbon as a result of carbon segregation through the metal surface.\cite{eizenberg}
Soon after, the surface segregation behavior of carbon from dilute solid solutions 
on Pt(100), Pt(111), Pd(100), Pd(111) and Co(0001) surfaces was investigated.\cite{hamilton}
Interestingly, much earlier it was argued that had the segregated carbon layer 
can be in the form of monolayer honeycomb structure like graphene.\cite{may}

Graphene,\cite{novo1} being not only the thinnest ever but also the strongest
material, has, in fact, the potential for nano-coating applications. When
sticks to or grown on various surfaces, graphene adds only negligible
thickness to the size of the underlying sample and forms an electrically
and thermally conductive coating on it. Moreover, graphene has exceptional 
mechanical, thermal and chemical stability. Various synthesis techniques of graphene 
covered metal surfaces and their electronic and structural properties have 
been reviewed by Winterlin \textit{et al.}\cite{winterlin} and Mattevi 
\textit{et al.}\cite{mattevi} Advances in the techniques of graphene synthesis have initiated 
the studies on graphene coating. Experimentally, Dedkov \textit{et al.}\cite{dedkov}
studied the oxygen protection of Fe intercalated Ni surface and bare Ni films. 
Borca \textit{et al.}\cite{borca} have experimentally demonstrated that the periodically
rippled structure of graphene can be grown on Ru(0001) surface and it serves as a perfect
coating material against oxidation. Much recently, Gadipelli \textit{et al.}\cite{taner} 
reported the formation of large-scale graphene monolayers on Cu surface, which is 
well-protected from the oxidation. Also the graphene coatings on Cu, Cu/Ni alloy, 
Pt and Ir surfaces have been exploited.\cite{coating1,coating2,coating3} XPS and SEM images 
presented evidences that Cu and Cu/Ni surfaces can be protected
from oxidation through graphene coating.\cite{coating1} However, despite these recent
progresses, very little is known how and why graphene layer constitutes a
protective coating on reactive surfaces and what are its limitations.

In this study we show that graphene can easily be oxidized by oxygen atoms which form strong
chemical bonds on its surface. Despite that the graphene coating can protect solid surfaces from
oxidation by posing a high energy barrier to any adsorbed oxygen atom diffusing from the top of
graphene to the interface between graphene and the reactive surface underneath. Because of this
barrier perpendicular diffusivity of oxygen atom is practically zero as compared to its lateral
diffusivity. Although an oxygen molecule is weakly bound to graphene and does not have any
direct role in the oxidation, it can be indirectly involved by dissociating into two atomic oxygens.
These oxygen atoms form relatively stronger chemical bonds with twofold coordinated carbons
but encounter much lower oxidation barrier when diffuse towards the reactive surface. Poor protection
from oxidation at defect sites can be circumvented by multilayer graphene coating.

\begin{figure}
\includegraphics[width=8cm]{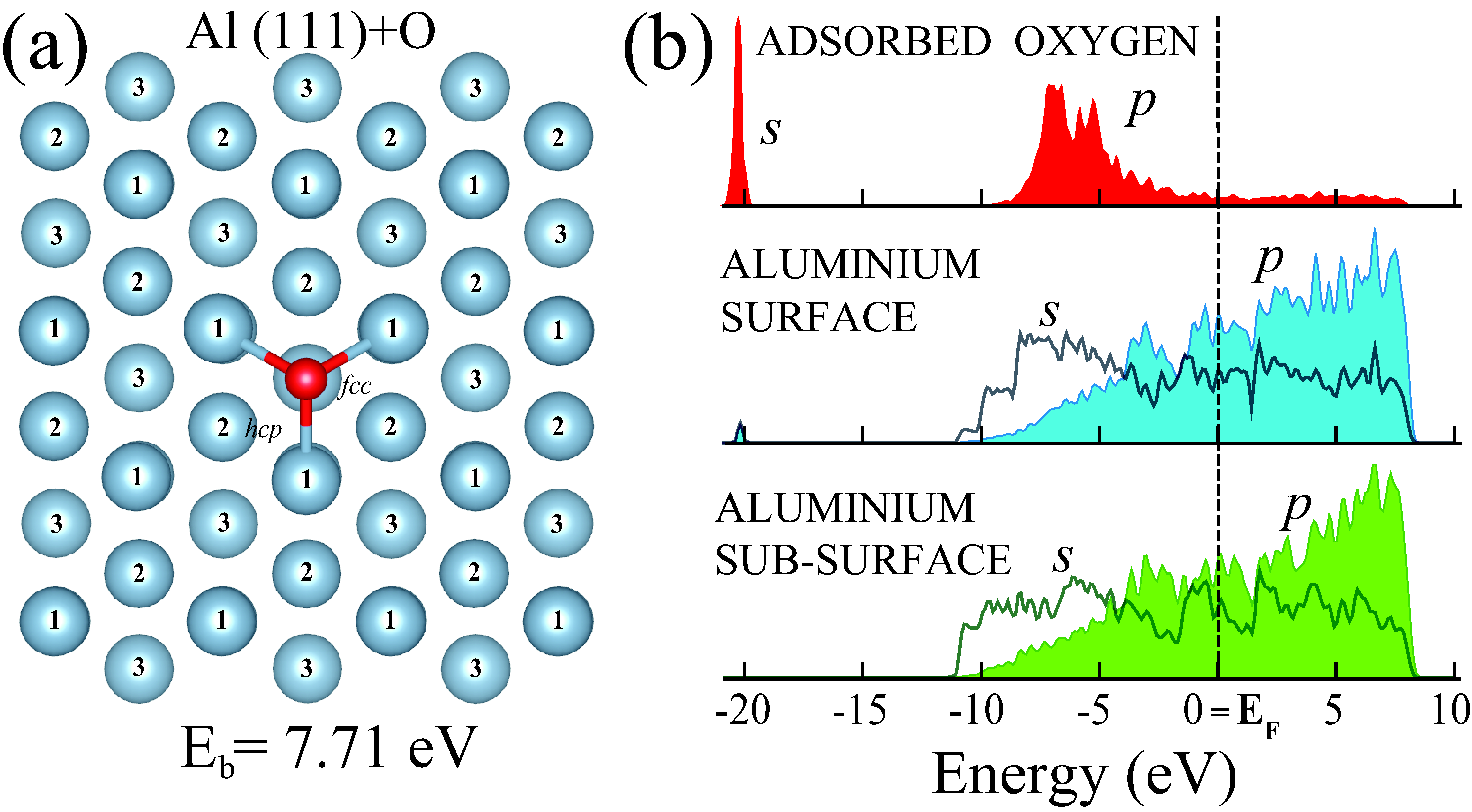}
\caption{(a) Atomic configuration of an oxygen atom adsorbed to Al(111) surface. Oxygen and Al 
atoms are illustrated by small-red and large-blue balls with numerals indicating their layer
numbers from the top. (b) Density of states (DOS) projected to $s$- and $p$-orbitals of 
adsorbed O, surface and subsurface layers of Al(111) slab.}

\label{f1}
\end{figure}

\section{Method}
Our study proceeds in three complementary and sequential steps: (i) In the first step we
examine the interaction of O$_2$ and O atoms with a bare reactive metal surface
and with a bare pristine graphene, where important features are discovered.
(ii) Second step deals with the sticking of graphene to a flat,
clean surface, which is vulnerable to oxidation. (iii) In the third step,
we show how graphene coating hinders oxygen atoms from diffusion towards the
protected surface. Our results are obtained by performing first-principles,
spin-polarized calculations within Density Functional Theory using VASP 
package.\cite{vasp1,vasp2} We used Generalized Gradient Approximation\cite{pbe} including
Van der Waals (vdW) correction\cite{grimme06}, PAW potentials\cite{paw} 
and a plane wave basis set with the kinetic energy cutoff of 500 eV. To minimize
the coupling between adsorbed atoms or molecules, the binding energies and reaction paths
are calculated using (4x4) or (6x6) supercells. For the coated surfaces a grid of
25x25x1 \textbf{k}-points is used. The convergence criterion of self consistent calculations
for ionic relaxations is taken $10^{-5}$ eV between two consecutive steps. By using the
conjugate gradient method, atomic positions and lattice constants are optimized until the
atomic forces are less than 0.05 eV/\AA. The energetics of various paths of O or O$_2$ are 
calculated by forcing them to pass through the graphene layer from above to below. The amount 
of displacement is identified as \textit{indentation} in the figures. The paths of minimum 
energy barrier are determined by relaxing carbon atoms of graphene, as well as lateral $x$- and 
$y$-coordinates of O or O$_2$ at each step of indentation corresponding to a fixed $z$-coordinate.
In some cases, the energy barriers associated with specific and well-determined paths
are also examined, where O$_2$ and O are forced to follow these paths, but the rest of 
atoms are relaxed. The maximum number of atoms treated in our calculations is 129, which
occurred in the determination of energy barriers associated with the coating of Al(111) surface 
by graphene bilayer.

\begin{figure}
\includegraphics[width=8cm]{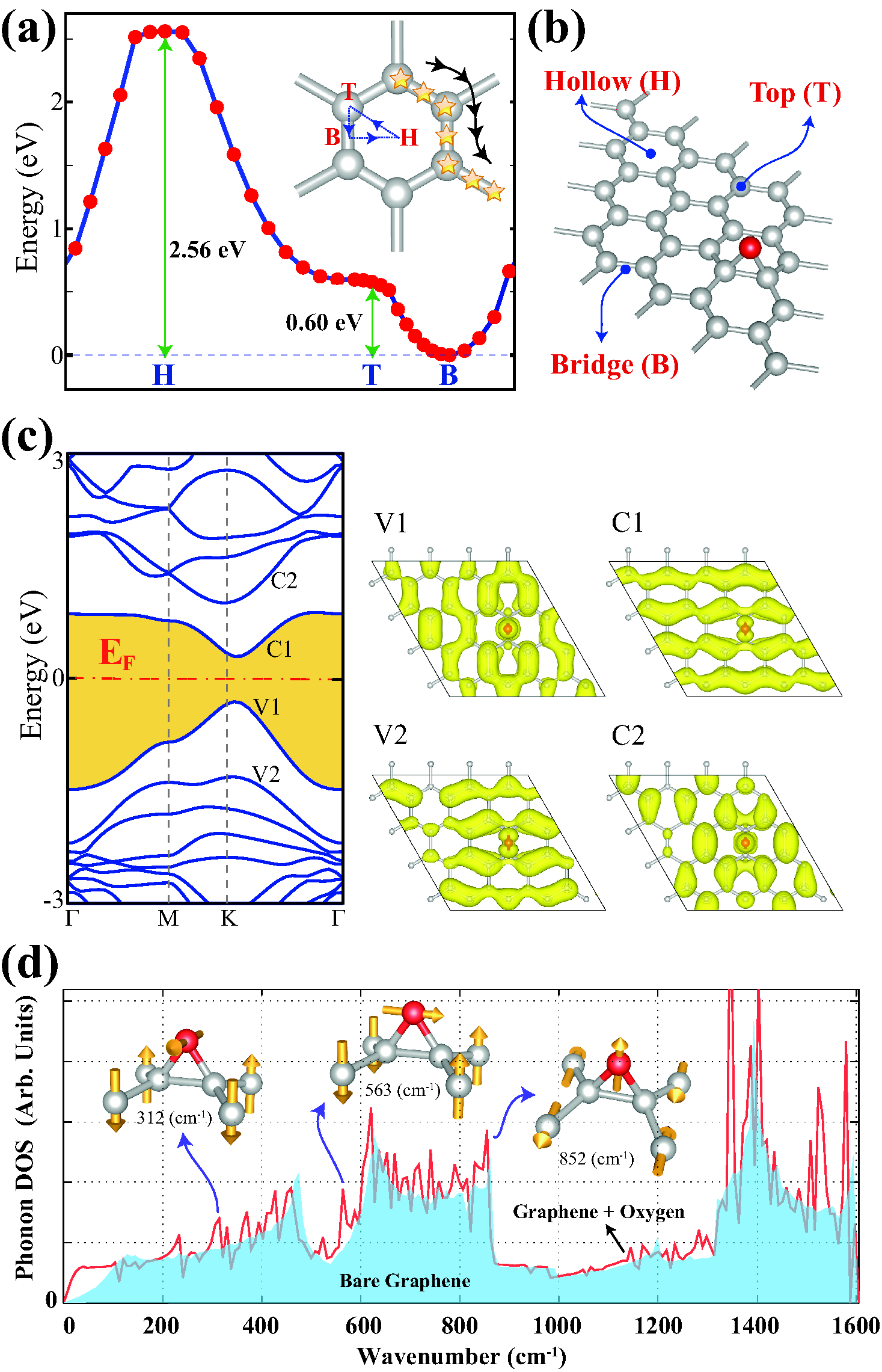}
\caption{(a) Variation of the energy of adsorbed oxygen atom along T(top)-H(hollow)-B(bridge) 
site directions of a hexagon showing that the B-site is energetically most favorable. Stars 
indicate favorable path for the diffusion of oxygen atom on the graphene surface. (b) Atomic 
configuration for an oxygen atom adsorbed at the bridge site on a (4x4) supercell of graphene 
consisting of 32 carbon atoms. (c) Electronic energy band structure together with the charge 
densities of specific conduction and valence band states. (d) Calculated density of phonon 
modes of a pristine graphene (shaded area) and those of oxygen adsorbed to the bridge site 
of the (4x4) supercell of graphene (red line). Relevant localized phonon modes are 
indicated by insets.}

\label{f2}
\end{figure}

\section{Oxidation of Al Surface and Graphene}
Since we are not concerned with sample specific details of oxidation behaviors of
the protected surfaces, Al(111) surface is taken here only as a prototype metal
surface vulnerable to oxidation when exposed to atmosphere, but is protected by
placing a graphene sheet between its surface and atmosphere. Al(111) surface alone
is represented by a 4-layer Al(111) slab as described in Fig.  \ref{f1}. It has
metallic and nonmagnetic ground state, and its states at the Fermi level ($E_F$) are 
composed of mainly $3p_{xy}$- and partially $3s$-orbitals of Al atoms. The work function
for this slab is calculated to be 4.06 eV, which is comparable to the value 
of 4.24 eV measured experimentally\cite{wf} for Al(111) surface. We calculate
that an oxygen atom is strongly bound to the Al(111) surface with 7.71 eV binding
energy at the fcc site and 7.24 eV at the hcp site. Fig. \ref{f1} (b) presents the
densities of electronic states projected to adsorbate O, surface and subsurface
layers of Al(111) slab. Apparently, $2p$-orbitals of adsorbed O mix with the $3p$
and $3s$-orbitals of Al substrate in a wide energy range to form a strong bond. Oxygen
molecule by itself interact strongly with Al(111) surface; it dissociates into
atomic oxygens which, in turn, are adsorbed at fcc and hcp sites.

\begin{figure}
\includegraphics[width=8cm]{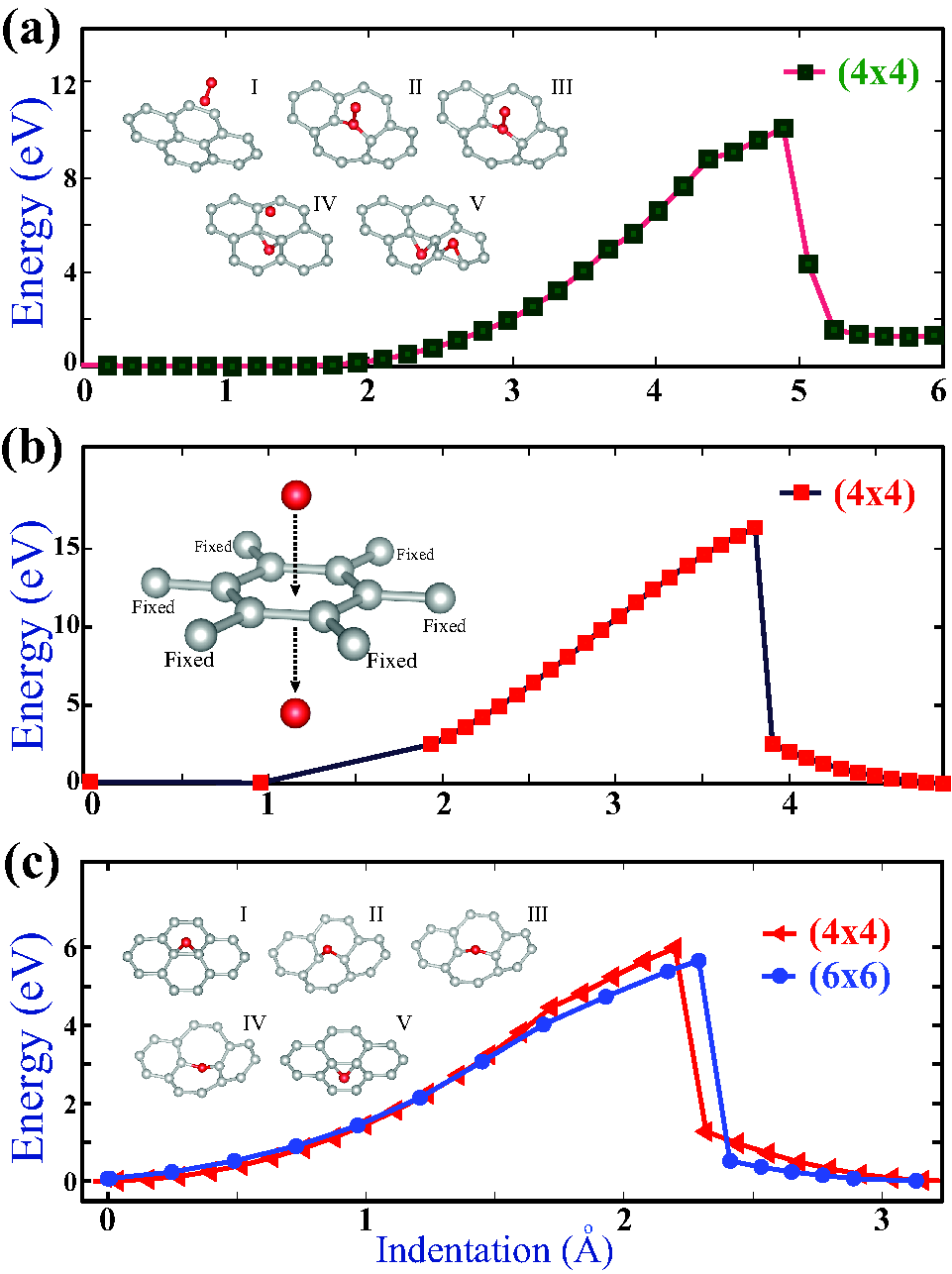}
\caption{Calculation of energy barriers of O$_2$ and O passing from the top to the 
bottom side of a suspended graphene along various paths. (a) O$_2$, which is forced 
to pass from the top to bottom side of graphene following a fixed vertical line 
through the hole at the center of a hexagon. (b) The energy barrier for O atom 
along the same path as (a). (c) The path of the minimum energy barrier for an O atom, 
which is initially adsorbed at the bridge site above the graphene plane is forced to pass to 
the bottom side. Positions of C atoms, as well as the lateral $x$- and $y$-coordinates 
of O are optimized for each value of indentation.}

\label{f3}
\end{figure}

Bare graphene can also be easily oxidized, whereby oxygen atoms are  adsorbed at the
bridge site positions above any C-C bond of graphene and become negatively charged. 
Our calculations using Bader analysis\cite{bader} estimates an excess charge of 0.79 
electrons at adsorbed O atom. The binding energy is calculated to be 2.72 eV at the 
bridge site, hence O atoms at the T(top)- or H(hollow)- (i.e. center of hexagon) site in 
Fig. \ref{f2} (a) move favorably to the B(bridge)-site. Figure \ref{f2} (c) presents the 
electronic  energy band structure corresponding to an O atom adsorbed to a bridge site 
of a (4x4) supercell of graphene and the charge density distributions of specific 
conduction and valence band states. Upon oxidation the linearly crossing $\pi$ and $\pi^*$
states of semimetallic bare graphene are modified and opened a band gap of 0.58 eV.
This explains why domains of dark (metallic) graphene surface becomes reflecting 
(insulator) upon oxidation.\cite{science} Reversible oxidation-deoxidation
of graphene through heating or charging has been pointed out as potential electronic
device application.\cite{science,dana} Reversibility is a strong evidence that graphene
surface remain chemically stable in the course of oxidation-deoxidation; neither bond
breaking nor modification of honeycomb structure does occur. However, the situation is
dramatically different for several other surfaces, such as Si, Fe, Te, Al, Cu etc,
where the chemical stability is destroyed upon oxidation.\cite{Fe, ciraci-Al, kiejna-Al,coating1}
Figure \ref{f2} (d) shows the density of phonon modes (DOS) of a pristine graphene 
and that of oxygen adsorbed to the bridge site of the (4x4) supercell of graphene, which
are calculated from the first principles.\cite{espresso} As seen, adsorbed oxygen 
atom gives rise to several localized phonon modes, which will be used in estimating 
the characterized frequency and the diffusivity thereof. 

In contrast to oxygen atom, an oxygen molecule has a weak binding with graphene. We calculated 
its binding energy to be 115 meV, and magnetic moment 1.90 $\mu_B$, slightly smaller than 
the magnetic moment of free O$_2$. Hence, an O$_2$ molecule with such a weak binding energy 
to bare graphene cannot have any significant effect on the oxidation of the protected surface, 
though the situation can be different for the defected graphene as we will show later. On the other hand,
a free O atom approaching another O atom already adsorbed on graphene forms a strong bond
with the latter and releases $\sim$ 4.13 eV energy in this exothermic process. Eventually an
O$_2$ molecule is formed thereof. This may explain why deoxidation of graphene is provided 
easily\cite{science} by the STM tip at 100$^\circ$ C at close proximity of oxidized graphene 
surface, despite the strong binding energy of O atom. Note that two adsorbed O at close proximity
can also form an O$_2$ molecule by releasing an energy of 1.60 eV, if they can overcome an energy barrier.

\section{Protection of Al(111) Surface by Graphene Coating}

Next we explore the protection of a reactive surface, such as Al(111) by sticking
or growing graphene on it. Sticking of graphene on various metals surfaces including
Al(111) surface has been studied earlier.\cite{graphene-coat-prb} Even though
Al(111) surface are not in registry with graphene honeycomb structure and has
hexagonal lattice constants $\sim$ 10\% larger than those of graphene, sticking of
graphene on this surface can be achieved. In order to present an estimation
value for the adsorption energy of graphene to Al(111) surface one has to compress
Al(111) slab laterally and expand graphene lattice in order to achieve the registry 
for the optimization of final structures using periodic boundary condition. Despite the
strain energy spent to obtain the lattice registry, the sticking occurs with
a significant binding energy of 2.67 eV per (4x4) supercell or ($\sim$166 meV per cell).
The sticking of graphene patches to the Al(111) surface in random orientation is a
complex and stochastic process, and can even lead to the formation of bubbles,
since low coordinated edge atoms have stronger binding with Al(111) surface.
Even if the average binding energy per carbon atom is small, it would require
significant energy to peal off the strong but flexible graphene layer from
the surface.

If graphene patches are placed randomly on Al(111), they may not be severely strained
to maintain the lattice registry. Therefore, we rather consider unstrained graphene and
compensate the lattice misfit by laterally compressing Al(111) slab. This way we achieve the
lattice registry to be able to use the periodic boundary conditions. Under these
circumstances, the adsorption energy of graphene to this compressed Al(111) slab,
which is  calculated to be 2.38 eV per (4x4) super cell (or $\sim$ 148 meV per
cell) is not significantly affected. Additionally, the binding energy of O atom
to the compressed Al(111) surface (7.15 eV) is still very high. Thus, despite the
compression dictated by the periodic boundary conditions, the compressed Al(111)
slab is still sufficiently reactive to mimic a surface to be protected
by graphene coating.

\subsection{Diffusion of O$_2$ and O through a suspended graphene}

Now we address the main issue pertaining to how a graphene coating, which by
itself is also vulnerable to oxidation can protect a reactive metal surface.
To clarify the mechanism of protection from oxidation we first examine how
an O$_2$ molecule or O atom can pass from one side of bare and suspended graphene to
the other side. The energetics and energy barriers involved in the course of these 
processes are calculated in a (4x4) supercell of bare graphene with specific carbon 
atoms are fixed to prevent the suspended layer from displacement. Here we consider 
first the fixed vertical path passing through the hole at the center of hexagons 
(as a seemingly possible diffusion path) and calculate the involved energy barrier 
when O$_2$ molecule or O atom are forced to follow this path as summarized in 
Fig. \ref{f3} (a) and (b). An oxygen molecule following this vertical path needs 
to overcome a barrier of 10.12 eV in Fig. \ref{f3} (a). However, once O$_2$ 
overcome this barrier, it dissociates into two O atoms; one O adsorbed above, 
the other adsorbed below at the bridge sites. Apparently, graphene acts as a 
membrane, that blocks the passage of O$_2$. If an O atom were forced to pass from the top
to the bottom side along this fixed vertical path, the energy barrier would be even 
higher, i.e. 16.34 eV in Fig. \ref{f3} (b). This fixed vertical path is not a possible
diffusion path with minimum energy barrier, since there are reaction paths with smaller 
energy barriers as explained below.

The path shown in  Fig. \ref{f3} (c) starts from the adsorbed oxygen atom at the 
bridge site at the top side, i.e. the minimum energy configuration of O atom adsorbed 
on pristine graphene. At each stage of indentation, carbon atoms, as well as $x$- 
and $y$-lateral coordinates of O atom are optimized to minimize the energy. As seen 
from the snapshots of atomic configuration corresponding to various stages, the 
passage of O takes place around the same bridge bond, whereby O atom switches from 
the top to the bottom side of graphene by gradually flattening C-O-C bridge bond.
The energy barrier to be overcame by an O atom to pass from the top to the bottom 
side is calculated to be $Q=$5.98 eV. To reveal whether the (4x4) supercell may 
impose constraints on the calculated energy barrier, we calculated the barrier 
in a relatively larger, (6x6) graphene supercell to be $Q=$5.65 eV. The calculated 
energy barrier is not affected from the size of the supercell and is high enough 
to block the diffusion and hence to hinder oxidation of the surface underneath. This
path is identified as the path of minimum energy barrier for O atom passing from 
the top to bottom of bare suspended graphene.

\begin{figure}
\includegraphics[width=7cm]{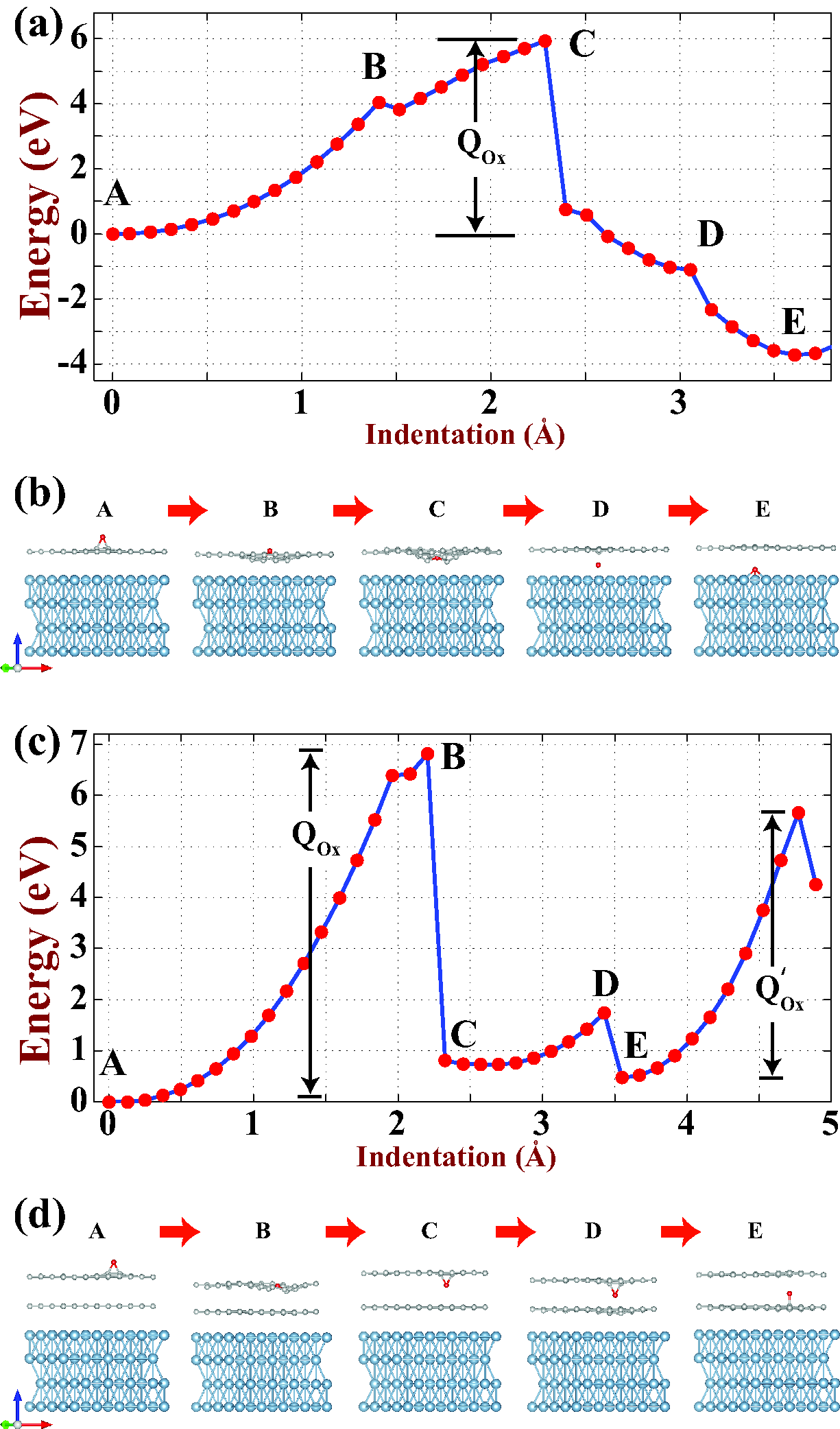}
\caption{ (a) Variation of the total energy for an O atom (red ball) passing (indenting) from the
top side of single graphene layer to its bottom side and eventually adsorbing to Al(111) surface
(blue balls) underneath. O atom follows the path of minimum energy barrier($Q_{ox}$=5.93 eV). (b) 
Snapshots of the atomic configurations corresponding to various stages between the initial stage A 
stating from the bridge site of graphene and final stage E ending with the adsorption of O atom on 
Al(111). (c) Protection of Al(111) surface from oxidation by a graphene bilayer and the variation 
of energy for an O atom penetrating from the top side of the outermost graphene layer and eventually 
adsorbing to Al(111). Highest barrier to be overcame by a diffusing O atom is $Q_{ox}=$6.81 eV along the
the path to reach to Al(111) surface. (d) Atomic configurations of various stages the case of (c).}

\label{f4}
\end{figure}

\subsection{Graphene Coating of Al(111)}
In the presence of the Al(111) slab underneath the protective graphene coating, we elaborate
and further optimize the reaction path in Fig. \ref{f3} (c) as the most likely pathway of oxidation.
The variation of the energy of an O atom moving along the reaction path of minimum barrier is shown in
Fig. \ref{f4} (a). The energy barrier along this reaction path is $Q_{ox}\sim$ 5.93 eV and
occurs as O is switching from the top side to the bottom side of graphene. Once the diffusing O atom
overcomes this barrier, it goes to Al(111) surface via the bridge site below without almost
any barrier and oxidizes the metal surface. This energy barrier $Q_{ox}$ is rather high 
and hence the protection of graphene against oxidation appears 
to be very efficient. The significance of high $Q_{ox}$ can be deduced by comparing the diffusivity  
of O moving on graphene surface $D_{\parallel}$ with that of O atom penetrating the graphene coating 
to oxidize the Al(111) surface, $D_{\perp}$. $D_{\parallel}=a^{2}\nu e^{Q/k_{B}T}$ can be estimated in 
terms of the energy barrier $Q=$0.60 eV in  Fig. \ref{f2} (a), lattice constant $a$=1.43 \AA~ and the 
characteristic vibration frequency extracted from the calculated localized phonon modes of O in   
Fig. \ref{f2} (d) to be $\nu \cong$ 22 THz. Accordingly, the diffusivity of O penetrating the graphene 
coating is estimated to be $D_{\perp}=D_{\parallel} \times 10^{-87}$, which is really negligible.

\begin{figure*}
\includegraphics[width=16 cm]{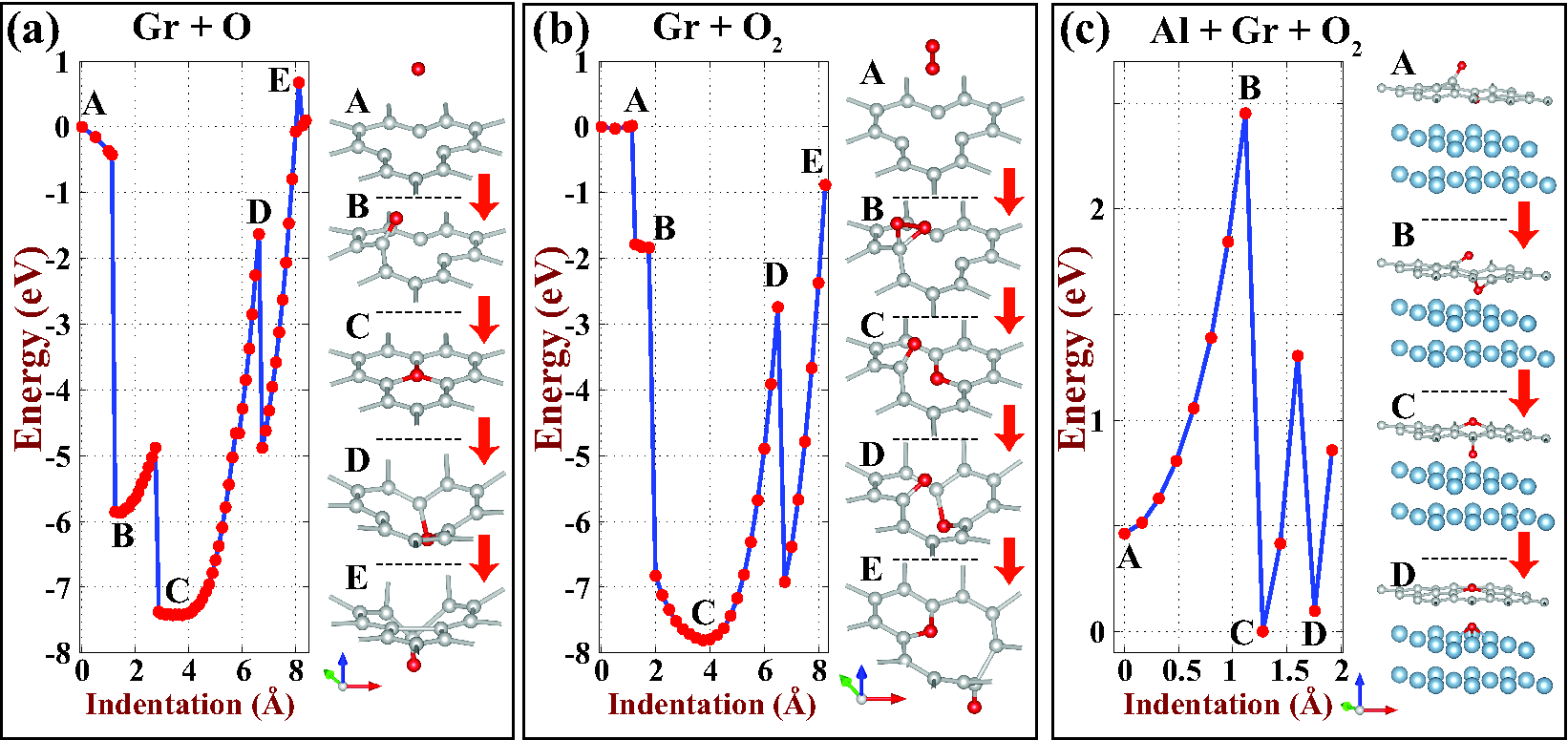}
\caption{(a) Evolutions of energetics and atomic structure with the indentation of
the oxygen atom, which is initially adsorbed at the edge of a single vacancy. (b) Evolutions of
energetics and atomic structure of two oxygen atoms resulting from the dissociation
of O$_2$ molecule at the edge of a single vacancy and the indentation of one of
oxygen atoms from the top to the bottom site of a suspended graphene. (c) The diffusion
of one of adsorbed oxygen atoms in (b) towards the Al(111) surface resulting in
its oxidation. Red, gray and blue balls indicate, respectively O, C and Al atoms.}

\label{f5}
\end{figure*}

\subsection{Bilayer Graphene Coating of Al(111)}
The effectiveness of the protection against oxidation can be further increased by coating with graphene
bilayer. In  Fig.  \ref{f4} (c) the variation of energy of diffusing O atom
from outermost graphene bilayer to the metal surface via second graphene layer following
the minimum energy reaction path is shown. Apparently, the oxidation barrier is
increased by 0.88 eV due to to the coating by graphene bilayer. Snapshots of relevant stages
in the course of diffusion of O starting above the first graphene layer through
the second graphene layer and eventually ending at the surface of Al(111) surface are also
indicated in  Fig.  \ref{f4} (d). While the oxidation barrier $Q_{ox}=$6.81 eV
occurs when the diffusing O switches from the top to the bottom side of first
graphene layer, there are additional barriers blocking the diffusion of O atoms.
For example, the energy barriers for switching from the bottom bridge site of first
layer graphene to the top bridge site of second graphene layer is $\sim$ 1 eV. If
this small barrier is overcame, the adsorbed O atom becomes attached to the second
graphene layer and is still separated from reactive surface. To proceed with
diffusion to reach the Al(111) surface one follows similar course as in  Fig.
\ref{f4} (a). Starting from the stage E, an O atom diffuses from the bridge 
site above the second graphene layer to the Al(111) surface by overcoming an energy
barrier of $Q^{'}_{ox}= 5.20$ eV and oxidizes the metal surface.

Clearly, the coating of reactive surfaces by sheets comprising more than two
layers of graphene will further increase the effectiveness of protection.
Sequential barriers posed at each graphene layer
increase the chance that the diffusing O can be trapped between any two barriers.
On the other hand, the size of the protected sample will be modified by each
additional graphene layer adding $\sim 3 \AA$ distance between O and Al surface, 
even if this increase of the thickness may be considered
negligible. Protection by graphene sheets comprising a few layers are
expected to be effective also to suppress the effects of any local heating or
energy transfer to the outermost graphene. For example, as pointed out at the
beginning a free oxygen at the close proximity of an adsorbed oxygen on graphene can
form O$_2$, whereby an energy of $\sim$ 4.13 eV is released. When deposited to the
graphene, this energy may create a local, nonequilibrium phonon distribution,
which was shown to dissipate within pico seconds.\cite{salim,haldun} Such a short
time interval is enough to accommodate several jumps of atoms. Hence a local
heating due to a chemical process may promote the diffusion of other adsorbed
oxygen atoms from the protective coating towards the reactive surface. Under these
circumstances while single layer graphene coating fails to hinder oxidation,
multilayer graphene coating can block the diffusing hot oxygen atoms.

\subsection{Vacancy Effect}
The above arguments related with the protection against oxidation relies on the
fact that graphene coating is continuous and defect free. If coating consists
of graphene patches, reactive surfaces cannot be covered at the zones between
patches, where they become directly exposed to atomic oxygens.\cite{coating1} The holes or
vacancies\cite{topsakal, iijima} of graphene are also spots, where oxygen atoms would penetrate
the metal surface without or relatively smaller energy barrier. In fact, the
etching of graphite following the dissociation of O$_2$ by producing CO and CO$_2$
have been pointed out.\cite{lee,scheffler} Here we consider the penetration of oxygen
atom near a vacancy in graphene. Three important features of our work are schematically
summarized in Fig. \ref{f5}. An oxygen atom can favorably bound to carbon atoms at the edge of
a single vacancy in Fig. \ref{f5} (a). The ground state is exothermic and releases
7.65 eV, whereby O atom substitutes the vacant carbon atom. In Fig. \ref{f5} (b)
an O$_2$ molecule can dissociate in two O atoms at the close proximity of a
vacancy in graphene. Subsequently, while one O atom is attached to one of three
twofold coordinated carbon atoms at the edge of the vacancy, the other one bridges between
the remaining two and hence completes the hexagon. In this exothermic process
7.83 eV energy is released in addition to the energy spent in dissociation process.
This shows that vacancies of graphene are active sites to catalyze the dissociation
of O$_2$ molecules. Fig. \ref{f5} (c) shows that the barrier in the diffusion
of a specific O atom adsorbed at the edge of a vacancy is dramatically
lowered ($Q_{ox} \sim$ 2 eV) and hence the protection from oxidation is weakened. Such a situation
shall occur at the grain boundaries and holes of graphene and confirms the
experiment\cite{coating1} that defects or discontinuities in covering a reactive
surface by graphene may result in the weakening of the oxidation protection. Similar
processes have been also confirmed at the edges of relatively larger holes. This serious
limitation caused by defects can be avoided by multilayer coatings.

\section{Conclusions}
In conclusion, we demonstrated that continuous coating of pristine graphene on reactive
surfaces can provide for an excellent protection from oxidation of reactive surfaces at
nanoscale. The binding of oxygen atom at low coordinated carbon atoms is rather high,
but their barrier to penetrate to the reactive surface under graphene is low. Therefore
discontinuities in graphene coating or defects, such as vacancies or holes
weaken the protection from oxidation by creating spots of low oxidation
barrier. This limitation can be circumvented by coating of bilayer or preferably graphene
sheets comprising a few graphene layers, which provides even more effective protection.
At macroscale, our results suggests that graphene additives can improve the strength
of antioxidant paints. Graphene coating, which is thin at the atomic scale can also serve
as a natural barrier between environment and solid surfaces of other elements.

This work is supported by TUBITAK through Grant No:108T234. All the
computational resources have been provided by TUBITAK ULAKBIM, High
Performance and Grid Computing Center (TR-Grid e-Infrastructure).
S. C. acknowledges the partial support of TUBA, Academy of Science of
Turkey.

\end{document}